\documentclass[11pt,a4paper]{article}
\date{}
\usepackage{amsmath}
\usepackage{float}
\usepackage{graphicx}
\usepackage[english]{babel}
\usepackage{amssymb}

\title{Casimir energy of an open string with angle-dependent boundary condition}

\title{Casimir energy of an open string with angle-dependent boundary condition}
\author{A. Jahan$^1$ and I. Brevik$^2$\\
$^1$Research Institute for Astronomy and Astrophysics of Maragha, Iran\\
$^2$Department of Energy and Process Engineering,\\ Norwegian University of Science and Technology 7491 Trondheim, Norway
\\\textit{jahan@riaam.ac.ir,\,iver.h.brevik@ntnu.no}}

\begin{document}
\maketitle

\begin{abstract}
We consider an open string with ends laying on the two different solid beams (rods). This set-up is equivalent to two scalar fields with a set of constraints at their end-points. We calculate the zero-point energy and the Casimir energy in three different ways: (1) by use of the Hurwitz zeta function, (2) by employing the  contour integration method in the complex frequency plane, and (3) by constructing the Green's function for the system. In the case of contour integration we also present a finite temperature expression for the Casimir energy, along with a convenient analytic approximation for high temperatures. The Casimir energy at zero temperature is found to be a sum of the L\"{u}scher potential energy and a term  depending on the angle between the beams.  The relationship of this model to an analogous open string model with charges fixed at its ends, moving in an electromagnetic field, is discussed.
\end{abstract}

\section{Introduction}
The Casimir energy is a physical manifestation of vacuum energy [1]. It is purely a quantum phenomenon which, for example, causes  two parallel conducting plates to attract each other. The vacuum energy of  open and closed strings, as  simple cases, has been investigated by several authors. L\"{u}scher \textit{et al} were the first ones who calculated the Casimir energy of an open string which is now called the L\"{u}scher potential [2-4]. They obtained this potential by considering a static quark-antiquark with the chromo-electric field between them as a vibrating string. The Casimir energy of a piecewise string was considered in [5-9]. The vacuum energy of an open string placed between two beads was obtained in [10]. The L\"{u}scher potential is recovered when the masses of the beads become large. The quantum corrections to the L\"{u}scher potential were calculated in [11], where the authors interpreted the corrections as a sort of non-local effect in a bosonic string. The Nambu-Goto model of an open string was used to model the inter-quark potential in [12, 13]. The string was assumed to end on point masses with mass $m$. It was shown that one recovers the L\"{u}scher potential in the limits $m\rightarrow 0, \infty$.  \\
The present work is to a large extent a sequel to a previous one where we obtained the Casimir energy as the zero-temperature limit of free energy of an open string in an angle-dependent set-up [14]. We there obtained the finite-temperature free energy using the path integral method. Here, we make use of three different methods to calculate the Casimir energy for  an open string whose ends are located on  two straight beams (solid rods). It is  assumed that there is  a relative angle $\theta$ between the beams, and as a result the boundary conditions for the ends of the string depend on $\theta$. We first derive the Casimir energy at $T=0$ by means of the Hurwitz zeta function, which under various circumstances has proved to be a very powerful method. Then, we employ the complex contour integration method, which gives a result necessitating numerical evaluation for general temperatures, but which nevertheless permits  a convenient analytical approximation  in the case of high $T$. We derive the angle-dependent Green's function, and calculate the Casimir energy on the basis of that. The energy turns out to be a  sum of an angle-dependent term and the L\"{u}scher potential. There are also interesting relationships between the present model and the one of an open string with charges at the ends, moving in an external electromagnetic field. This issue is discussed in the Summary section.

\section{Classical Dynamics}
 Assume a string which its ends lie upon two beams and can freely slide on them. The string has tension $T$ and mass density $\mu$. One of the beams is located at $z=0$ and the other at $z=l$. The angles between the $X$-axis and the rods at $z=0$ and $z=l$ are $\theta_1$ and $\theta_2$, respectively. The displacement of the string from equilibrium is parallel to the $X-Y$ plane and can be described by a displacement field  $\boldsymbol\phi(z,t)$, which is written as
\begin{equation}
\boldsymbol{\phi}(z,t)={\phi}_1(z,t)\widehat{e}_1+{\phi}_2(z,t)\widehat{e}_2.
\end{equation}
 Here $\widehat{e}_1$ and $\widehat{e}_2$ are  unit vectors along the $X$-axis and $Y$-axis, respectively. The Lagrangian density for the displacement field is given by
\begin{equation}\label{3}
\mathcal L=\frac{1}{2}\mu A\bigg(\frac{\partial\boldsymbol{\phi}(z,t)}{\partial t}\bigg)^2-\frac{1}{2}T\bigg(\frac{\partial\boldsymbol{\phi}(z,t)}{\partial z}\bigg)^2,
\end{equation}
which yields the the wave-equation
\begin{equation}\label{1a}
\bigg(\frac{1}{v^2}\frac{\partial^2}{\partial t^2}-\frac{\partial^2}{\partial z^2}\bigg)\boldsymbol{\phi}(z,t)=0,
\end{equation}
where the speed of sound is $v=\big(\frac{T}{\rho}\big)^{\frac{1}{2}}$. The ends of the string satisfy the constraints
\begin{equation}
\phi_2(0,t)-\tan\theta_1 {\phi}_1(0,t)=0, \label{constraints1}
\end{equation}
\begin{equation}
\phi_2(l,t)-\tan\theta_2 \phi_1(l,t)=0. \label{constraints2}
\end{equation}
These constraints lead to the following boundary conditions at the ends of string,
{\setlength\arraycolsep{2pt}
\begin{eqnarray}
\frac{\partial\phi_1(0,t)}{\partial z}+\tan\theta_1\frac{\partial\phi_2(0,t)}{\partial z}&=&0,\\
\frac{\partial\phi_1(l,t)}{\partial z}+\tan\theta_2\frac{\partial\phi_2(l,t)}{\partial z}&=&0.
\end{eqnarray}}
Then the solutions become [14]
{\setlength\arraycolsep{2pt}
\begin{eqnarray}
\phi_1(z,t)&=&\sum_{n=-\infty}^\infty \Big(a_ne^{i\omega_nt}+\bar{a}_ne^{-i\omega_nt}\Big)\cos (k_nz+\theta_1),  \\
\phi_2(z,t)&=&\sum_{n=-\infty}^\infty \Big(a_ne^{i\omega_nt}+\bar{a}_ne^{-i\omega_nt}\Big)\sin (k_nz+\theta_1),
\end{eqnarray}
where  the quantized wave numbers $k_n$ are
\begin{equation}\label{1}
k_n=\frac{\pi}{l}(n+r),\quad\quad r=\frac{1}{\pi}(\theta_2-\theta_1),
\end{equation}
and the corresponding eigenfrequencies
\begin{equation}
\omega_n=\frac{\pi v}{l}(n+r).\label{eigenfrequencies}
\end{equation}

As is  seen from (3), (4) and (5), one can interpret the whole set-up as a free field theory with a set of constraints at the end points $z=0, l$ of the string. We put in the next section $\theta_1=0$ and $\theta_2=\theta$, so that $r=\theta/\pi$. In section 4, we set again $\theta=\theta_2-\theta_1$ with $\theta_1\neq 0$.

The basic dispersion relation is thus
\begin{equation}
\sin \frac{\omega l}{v}-\tan \theta \cos \frac{\omega l}{v}=0. \label{dispersion}
\end{equation}
It may be of interest to note that this is formally the same kind of dispersion relation as encountered in solid state physics, in the so-called  Kronig-Penney model, in the degenerate case when the contribution from the quasi-momentum (defining the Bloch periodicity) is zero \cite{kronig31}.

\section{Casimir energy at zero and at finite temperature}

To evaluate  the Casimir energy connected with the eigenfrequencies (\ref{eigenfrequencies}), we shall make use of two different methods. They both prove to be elegant and  effective, and are reasonably easy to implement. First, we focus on the zero temperature case, $T=0$.

\subsection{Use of the Hurwitz zeta function}

General treatises on this regularization method can be found in Refs.~\cite{elizalde94} or \cite{elizalde95}. (The first application to the analogous composite string system was made by Li {\it et al.} \cite{li91}.) The Hurwitz function $\zeta_H(s,a)$ is originally defined as
\begin{equation}
\zeta_H(s,a)=\sum_{n=0}^\infty (n+a)^{-s}, \quad (0<a<1, \quad \Re{s}>1). \label{Hurwitz}
\end{equation}
The Hurwitz function in this form is defined only for $\Re{s}>1$; it is a meromorphic function with a simple pole in $s=1$. For $\Re{s}<1$ it can be analytically continued in the complex plane. In practice, one needs actually only the following property of the analytically continued  function, corresponding to $s=-1$,
\begin{equation}
\zeta_H(-1,a)=-\frac{1}{2}\left( a^2-a+\frac{1}{6}\right). \label{x}
\end{equation}
We assume $0 \leq \theta \leq \pi$, so that $0 \leq r \leq 1$. In the expression (\ref{eigenfrequencies}) for the eigenfrequencies we include only the positive values of $\omega_n$; it means counting from $n=0$ upwards. (This is related to our description of the waves as standing waves. If propagating modes were considered as the basic modes instead, the left-moving waves would be associated with the negative values of $n$.) The expression for the total zero-point energy $E_0$ becomes then, in unregularized form,
\begin{equation}
E_0=\frac{\pi v}{2l}\sum_{n=0}^\infty (n+r),
\end{equation}
which can be further processed using Eq.~(\ref{x})
\begin{equation}
E_0=\frac{\pi v}{2l}\zeta_H(-1,r) = -\frac{\pi v}{4l}\left( r^2-r+\frac{1}{6}\right). \label{6}
\end{equation}
To obtain the Casimir energy $E_C$, we have to subtract off a counter term $E_{\rm counter}$, corresponding to the zero-point energy in the case of zero deflection $\theta=0$,
\begin{equation}
E_{\rm counter}=\frac{\pi v}{2l}\zeta_H(-1,0)=-\frac{\pi v}{24l}.
\end{equation}
The final answer becomes thus, at $T=0$,
\begin{equation}
E_C=E_0-E_{\rm counter}=\frac{\pi v}{4l}r(1-r). \label{6a}
\end{equation}
We have $E_C=0$ for $\theta=0$ as it should according to construction, and the same answer follows also for $\theta=\pi$. The maximum value is obtained for $\theta=\frac{\pi}{2}$,
\begin{equation}
E_C\big|_{\rm max}=\frac{\pi v}{16l}.
\end{equation}

\noindent A characteristic property of $E_C$ is that it is non-negative. This contrasts the behavior found for the Casimir energy in most systems, as this  energy is usually negative, corresponding to an attractive Casimir force between two parallel plates as a typical example. What is the physical reason for the positivity of $E_C$ in the present case? We suggest that the reason is that the establishment of the final configuration  $\theta >0$ from the initial state $\theta=0$ requires a  {\it work} done on  the system from the outside. This means an increase of the system's mechanical energy, inducing in turn an increase of its zero-point energy.

\noindent To put this result into a broader perspective, let us compare with the Casimir theory for a piecewise uniform string. The original theory for such a system was developed in [7], for $T=0$.  The model was that of a closed string of total length $l=l_I+l_{II}$ consisting of two pieces $l_I$ and $l_{II}$,  subject to two boundary conditions at the junctions: (i) continuity of the transverse displacements, and (ii) continuity of the transverse elastic forces. The model was relativistic, in the sense that the velocity of sound was assumed to be the velocity of light in both of the pieces. With the tension ration defined as $x=T_I/T_{II}$ and the length ratio defined as $s=l_{II}/l_I$  the dispersion equation turned out to be
\begin{equation}
\frac{4x}{(1-x)^2}\sin \frac{\omega \pi}{2}+\sin \left(\frac{\omega \pi}{1+s}\right) \sin \left( \frac{\omega s\pi}{1+s}\right)=0.
\end{equation}
This equation can be solved by various methods. For simplicity we restrict ourselves here to the case of a small tension ratio, $x\rightarrow 0$. Then the eigenvalue spectrum for the two branches becomes distributed over two sequences
\begin{equation}
\omega_n(s)=(1+s)\pi n/l, \label{string1}
\end{equation}
\begin{equation}
\omega_n(s^{-1})=(1+s^{-1})\pi n/l. \label{string2}
\end{equation}
From these expressions the contrast to our present model is evident: the eigenfrequencies in (\ref{string1}) and (\ref{string2}) are {\it proportional to} $n$. There is no property of  the composite string model that causes the eigenvalue equation to be inhomogeneous like (\ref{eigenfrequencies}). \\
\noindent Another related case of interest to compare with, is the so-called quantum spring \cite{feng10}. This system considers the oscillations of a massless scalar field under helix boundary conditions, and the Casimir force parallel to the axis of the helix is similar  to the elastic force in a spring. In this case the eigenvalues turn actually out to be inhomogenous in the integer number $n$, the inhomogeneity arising from the pitch of the circumference of the helix.

\subsection{Contour integration method}

The sum over $n$ in the evaluation of the zero-point energy can be expressed as a contour integral, exploiting that any meromorphic function $g(\omega)$ satisfies the equation
\begin{equation}
\frac{1}{2\pi i}\oint \omega \frac{d}{d\omega}\ln g(\omega)d\omega= \sum \omega_0 -\sum \omega_\infty,
\end{equation}
where $\omega_0$ means the zeros and $\omega_\infty$ the poles of $g(\omega)$ inside the integration contour. This contour will be chosen to be a semicircle of large radius $R$ in the right half of the $\omega$ plane, closed by a straight line from $\omega=iR$ to $\omega =-iR$. This procedure, usually called the argument principle, is treated in some detail by Barash {\it et al.} [20], for example. In connection with Casimir theory, the principle was introduced by van Kampen {\it et al.} [21].

To begin with, consider the following ansatz for $g(\omega)$:
\begin{equation}
g(\omega)=\Big|\sin \frac{\omega l}{v}-\tan \theta \cos \frac{\omega l}{v}\Big|^2.
\end{equation}
This function has correct zeros on the real axis, and has no poles. The divergence encountered when summing over all frequencies can be avoided formally by introducing a convergence factor $e^{-\alpha \omega}$, with $\alpha$ a small positive parameter. Moreover, when aiming at calculating the zero point energy caused by nonzero values of $\theta$, we divide by the factor $\sin^2(\omega l/v)$ so that the argument of the logarithm becomes equal  to one when $\theta=0$.  Finally, we divide with the constant factor $(1+\tan^2\theta)$, for reasons to be clear below. Thus,
\begin{equation}
g(\omega) \rightarrow \bigg| \frac{\sin \frac{\omega l}{v}-\tan \theta \cos \frac{\omega l}{v}}
{\sin{\frac{\omega l}{v}}}\bigg |^2 \frac{1}{1+\tan^2\theta}.
\end{equation}

 On the imaginary axis where $\omega=i\xi$, one has
\begin{equation}
g(i\xi)=\frac{1+   \tan^2\theta\coth^2 \frac{\xi l}{v}}{1+\tan^2 \theta},
\end{equation}
showing that $g(i\xi)\rightarrow 1$ when $\xi \rightarrow \pm \infty$. These extremal points, together with the other points on the big semicircle, do not contribute. We can thus write, for the Casimir energy at $T= 0$,
\begin{equation}
E_C=-\frac{1}{2\pi}\int_0^\infty\xi\frac{d}{d\xi}\ln g(i\xi)d\xi.
\end{equation}
We here make a partial integration, observing that the boundary terms for $\xi=0$ and $\xi=\infty$ do not contribute. The  Casimir energy then  becomes, when we finally insert the convergence factor,
\begin{equation}
E_C=\frac{1}{2\pi}\int_0^\infty e^{-i\alpha \xi}\ln\left( \frac{1+   \tan^2\theta \coth^2\frac{\xi l}{v}}{1+\tan^2\theta}\right) d\xi.
\end{equation}
An advantage of the contour integration method is that the expression can easily be generalized to the case of finite temperatures. The general substitution is
\begin{equation}
\hbar \int_0^\infty d\xi \rightarrow 2\pi k_BT{\sum_{n=0}^\infty}^\prime
\end{equation}
(here expressed in dimensional units), where the prime means that the $n=0$ term is taken with half weight. The discrete Matsubara frequencies are $\xi_n=2\pi nk_BT/\hbar$, where $n=0,1,2,..$. The Casimir free energy at finite $T$ becomes accordingly
\begin{equation}
F_C(T)=k_BT\sum_{n=1}^\infty e^{-i\alpha \xi_n}\ln \left( \frac{1+  \tan^2\theta \coth^2 \frac{\zeta_nl}{v} }{1+\tan^2\theta}\right),\label{finitetemperature}
\end{equation}
where we let the summation start from $n=1$ to avoid the divergence for $n=0$. Again, by construction, $F_C(T)$ becomes zero when $\theta=0$. One needs to distinguish between high and low temperatures.  There are two natural frequencies here; the first is the thermal frequency $\omega_T=k_BT/\hbar$; the second is the geometric frequency $\omega_{\rm geom}=2\pi c/l$ associated with the size of the system. A high-temperature state is characterized by the frequency ratio $\omega_T/\omega_{\rm geom}$ being large,
\begin{equation}
\frac{\omega_T}{\omega_{\rm geom}}=\frac{k_BTl}{2\pi \hbar c}=\frac{\xi_1l}{(2\pi)^2\hbar c} \gg 1.
\end{equation}
In the low-temperature state, this parameter is small.
\subsubsection{High temperatures}
For low temperatures the expression (\ref{finitetemperature}) is complicated, due to the large variation of the coth function in the frequency region $n \in [1, \infty]$. One can process it using the Euler-Maclaurin formula or the Abel-Plana formula, but we shall henceforth limit ourselves to high temperatures only. This case is easy to analyze, and it is moreover able to demonstrate the characteristic properties of our mechanical system.
  Using that $\coth z \approx 1+2e^{-2z}$ when $z\gg 1$, we see that  the argument of the logarithm in (\ref{finitetemperature}) can be replaced by $1+4\sin^2 \theta e^{-2z}$ where $z=\xi_nl/v$. The main contribution occurs for the lowest mode, $n=1$. Thus we obtain the following high-temperature expression for the free energy ($\hbar=1$)
\begin{equation}
F_C(T)= 4k_BT\sin^2 \theta e^{-4k_BTl/v} \label{highT}
\end{equation}
(the cutoff factor does not play a role at high temperatures).

It is instructive to compare this with  the expression for the high-temperature Casimir free energy for a pair of conducting plates separated by a gap $a$ [1],
\begin{equation}
F_C(plates)=-\frac{k_BT}{8\pi a^2} \zeta(3)-\frac{k_BT}{4\pi a^2}e^{-4\pi  k_BTa}. \label{elmag}
\end{equation}
The following observations can be made:
\begin{enumerate}
\item The first and dominant term in (\ref{elmag}), the term proportional to $T$, is lacking in (\ref{highT}). This is the term corresponding to $n=0$ in the electromagnetic case, and corresponds to classical theory. In the present model, there is no such particular role played by the case $n=0$.

\item The first term in (\ref{highT}) and the second term in (\ref{elmag}) are of the same kind, as they contain $T$ multiplied with a decreasing exponential in $T$. In the exponentials, $l/v$ corresponds to $\pi a$.

\item The maximum free energy in (\ref{highT}), for fixed $T$, occurs when $\theta =\frac{1}{2}\pi$. This is the same behaviour as we found above for $T=0$.

\item Finally, the expression (\ref{highT}) and the second term in (\ref{elmag}) have opposite signs. We commented on this point already above. It can be further illustrated by the following argument: Assume that our string system is slowly displaced from $\beta$ to $\beta +d\beta$, at constant temperature. This process requires positive external work, and the Casimir free energy increases. In the electromagnetic case, if the plates are displaced from $a$ to $a+da$, the process also requires positive external work. In that sense, the two cases are parallel to each other. The difference lies in that the string system approaches a maximum energy state at $\theta=\frac{1}{2}\pi$, while the plate system approaches the case $a=\infty$ where the free energy is {\it zero}. In turn, this gives rise to different signs in the Casimir free energies.
\end{enumerate}

\subsubsection{Other thermodynamic potentials}

It is of interest to calculate other thermodynamic potentials also, still assuming high temperature. Thus it easy to calculate the system's Casimir entropy $S_C$, using the general formula $S=-\partial F/\partial T$. We obtain
\begin{equation}
S_C=-4k_B \sin^2\theta \left( 1 - \frac{4 k_BTl}{v}\right)  e^{-4k_BTl/v}.
\end{equation}
This expression can have either sign, depending on the magnitude of $4k_BTl $ versus $v$. It is well known from other cases - cf., for instance, Refs.~\cite{milton19} and \cite{hoye16} - that  Casimir entropies can be negative. This implies no conflict with the second law in thermodynamics   as we are dealing with  a part of the complete system only -  the second law applies to the complete system.  For moderately high temperatures, $S_C<0$. If $4k_BTl/v >1$ then $S_C$ becomes positive, but its magnitude is suppressed by the exponential. For $T\rightarrow \infty$, $S_C \rightarrow 0$.

The internal Casimir energy $U_C$ can be found using $F=U-TS$. We obtain
\begin{equation}
U_C= 4k_BT\left(\frac{4k_BTl}{v}\right) \sin^2\theta e^{-4k_BTl/v}.
\end{equation}
This expression is always positive.

\section{Field theoretical approach: Green's function}

We consider now the problem from a field theoretical point of view, where a central point is to determine the Green's function. In this section we take into account the whole frequency region including negative frequencies, so that $n \in [-\infty, \infty]$. The eigenmodes, defined as
\begin{eqnarray}
u^1_n(z)&=&\frac{1}{\sqrt{l\,}}\cos (k_nz+\theta_1),\\
u^2_n(z)&=&\frac{1}{\sqrt{l\,}}\sin (k_nz+\theta_1),
\end{eqnarray}
satisfy the orthogonality relations
{\setlength\arraycolsep{2pt}
\begin{eqnarray}
\nonumber
\int _0^ldz\,\sum_{\alpha=1}^2u^\alpha_{n'}(z)u^\alpha_n(z)&=&\frac{1}{l}\int _0^ldz\cos z(k_n-k_{n'})\\
&=&\delta_{n'n}, \label{33}
\end{eqnarray}}
and fulfill the closure relation
\begin{eqnarray}
\nonumber
\sum_{n=-\infty}^\infty u^\alpha_{n}(z)u^\beta_n(z')&=&K^{\alpha\beta}(z,z')\delta(z-z').\\
&\equiv&\delta^{\alpha\beta}\delta(z-z')\label{38}
\end{eqnarray}
The explicit form of the matrix $K^{\alpha\beta}(z,z')$ is
\begin{equation}
K^{\alpha\beta}=
\left[ \begin{array}{ccc}
\cos\frac{\pi r}{l}(z-z') &-\sin\frac{\pi r}{l} (z-z') \\
\sin\frac{\pi r}{l}(z-z') &\quad\cos\frac{\pi r}{l}(z-z') \label{40}
\end{array} \right],
\end{equation}
which can be obtained with the help of the following Fourier series representations
{\setlength\arraycolsep{2pt}
\begin{eqnarray}
\frac{1}{l}\sum_{n=-\infty}^\infty \cos\frac{\pi nz}{l}\cos\frac{\pi nz'}{l}&=&\delta(z-z'),  \label{41}\\
\frac{1}{l}\sum_{n=-\infty}^\infty \sin\frac{\pi nz}{l}\sin\frac{\pi nz'}{l}&=&\delta(z-z'). \label{42}
\end{eqnarray}}
The second equality in (39) can be understood by noting that $K^{\alpha\beta}(z,z')|_{z=z'}=\delta^{\alpha\beta}$. Inspection of (\ref{40}) also shows that $K^{\alpha\beta}(z,z')|_{r=0}=\delta^{\alpha\beta}$, which implies that the Green's function is diagonal when the beams are parallel. The Green's function $G^{\alpha\beta}$ can be expanded in terms of the eigenmodes (36) and (37) as
\begin{equation}
G^{\alpha\beta}(t-t';z,z')=\int_{-\infty}^\infty\frac{d\omega}{2\pi}e^{i\omega(t-t')}g^{\alpha\beta}(\omega;z,z'),\label{44}
\end{equation}
with
\begin{equation}
g^{\alpha\beta}(\omega;z,z')=\sum_{n=-\infty}^\infty\frac{u^\alpha_{n}(z)u^\beta_n(z')}{\lambda_n(\omega)}, \label{45}
\end{equation}
where the eigenvalues are
\begin{equation}
\lambda_n(\omega)=k^2_n-\frac{\omega^2}{v^2}. \label{46}
\end{equation}
With the aid of (\ref{38}), (43) and (44) we obtain
\begin{equation}
\bigg(\frac{1}{v^2}\frac{\partial^2}{\partial t^2}-\frac{\partial^2}{\partial z^2}\bigg)G^{\alpha\beta}(t-t';z,z')=\delta^{\alpha\beta}\delta(t-t')\delta(z-z'). \label{47}
\end{equation}
\section{Alternative derivation of the Casimir energy at $T=0$}

The zero-point energy in terms of the Green function is given by [1]
\begin{equation}
E_0=\frac{i}{2\mathcal T}\textrm{Tr}\ln G^{\alpha\beta}(t-t';z,z'). \label{52}
\end{equation}
where $\mathcal T$ is (infinite) time interval. For the Green's function (\ref{44}), with the help of (\ref{45}) and (\ref{33}) we obtain the zero-point  energy as
\begin{equation}
E_0=\frac{1}{2i}\int_{-\infty}^\infty\frac{d\omega}{2\pi}\sum_{n=-\infty}^\infty\ln\bigg(k^2_n-\frac{\omega^2}{v^2}\bigg). \label{54}
\end{equation}
By performing an Euclidean rotation $\omega\rightarrow i\xi$ and using the Riemann zeta-function regularization in a straightforward way
{\setlength\arraycolsep{2pt}
\begin{eqnarray}
\nonumber
\sum_{n=-\infty}^\infty 1&=&1+2\zeta(0)\\ \label{55}
&=&0,
\end{eqnarray}}
we recast (\ref{54}) into (see appendix A)
{\setlength\arraycolsep{2pt}
\begin{eqnarray}
\nonumber
E_0&=&\frac{v}{2l}\int_{0}^\infty d\kappa\sum_{n=-\infty}^\infty\ln\bigg[\Big(n+\frac{\theta}{\pi}\Big)^2+\kappa^2\bigg]\\\nonumber
&=&\frac{v}{2l}\int_{0}^\infty d\kappa\Big[\ln (1-e^{-2\pi\kappa-2i\theta})\\
&& +\ln(1-e^{-2\pi\kappa+2i\theta})\Big], \label{56}
\end{eqnarray}}
where $\kappa=\frac{\omega l}{\pi v}$. Here we have neglected the irrelevant quadratically divergent terms in second line of (\ref{56}). Now, to evaluate the last integrals of (\ref{56}) we proceed by engaging the the series expansion of the logarithm for a given complex number $Z$
\begin{equation}
\ln(1-Z)=-\sum_{m=1}^\infty\frac{Z^m}{m}. \label{57}
\end{equation}
provided that $|Z|< 1$. So, we can write
{\setlength\arraycolsep{2pt}
\begin{eqnarray}
\int_{0}^\infty d\kappa\ln (1-e^{\pm 2i\theta}e^{-2\pi\kappa})&=&-\sum_{m=1}^\infty\frac{e^{\pm 2im\theta}}{m^2}.
\label{58}
\end{eqnarray}}
We substitute (\ref{58}) in (\ref{56}) and use the identity [24]
\begin{equation}
\sum_{m=1}^\infty\frac{\cos m x}{m^2}=\frac{\pi^2}{6}-\frac{\pi x}{2}+\frac{x^2}{4},
\end{equation}
which results in the zero-point energy as
{\setlength\arraycolsep{2pt}
\begin{eqnarray}
\nonumber
E_0&=&-\sum_{m=1}^\infty\frac{\cos 2m\theta}{m^2}\\
&=&-\frac{\pi v}{2l}\left(r^2-r+\frac{1}{6}\right)         \label{60}
\end{eqnarray}}
(recall $r=\frac{\theta}{\pi}$). Subtracting off the counter term $E_{\rm counter}$ corresponding to $r=0$ we thus get for the Casimir energy
\begin{equation}
E_C=E_0-E_{\rm counter}= \frac{\pi v}{2l}r(1-r). \label{61}
\end{equation}
 These results are in agreement with those obtained earlier in Sec.~3 (the reason why the expressions (\ref{60}) and (\ref{61}) are twice the corresponding expressions (\ref{6}) and (\ref{6a}) is that we have in the present section included the whole span $n\in [-\infty,\infty]$).

 From a field theoretical perspective, Eq.~(\ref{60}) yields the L\"{u}scher potential when the rods are parallel ($\theta=0$) or anti-parallel ($\theta=\pi$). In these two cases the string satisfies the Neumann-Neumann and Dirichlet-Dirichlet boundary conditions. For $\theta=\frac{\pi}{2}$, the ends of string obey the Neumann-Dirichlet boundary condition and the zero-point  energy  (\ref{60}) raises to its maximum value $E_{0}|_{\rm max}=\frac{1}{24}\frac{\pi v}{l}$ [14].

 \section{ Conclusions and final remarks}
 Casimir theory for string system occurs in various variants. As mentioned above, in Section 3, there is a  relationship of the present theory with the theory of a piecewise uniform string. Another variant that we wish to elaborate on somewhat further is the theory of a string having electric charges at its ends, and is interacting  with an external electromagnetic field. The theory for that kind of string was developed by Nesterenko \cite{nesterenko89}. It is of interest to elucidate the similarities  and the differences between that electromagnetic string theory and the present one. First, it turns out that the governing wave equations are basically the same; cf. our wave equation (\ref{1a}) above. Second, the   difference turns out to lie  in the boundary conditions.  In the electromagnetic model the  string coordinates are called $ {x^\mu} (\tau, \sigma)$ in \cite{nesterenko89}, where
  $\tau $ is the time coordinate and $\sigma$ the length coordinate along the string, and the boundary conditions are that the positions of the ends are kept at  rest, at $\sigma=0$ and $\sigma=\pi$. This means that there is a balance between the elastic force $Tx_\mu'$ and the electromagnetic force $qF_{\mu\nu}(x){\dot{x}}^\nu$ at each end point of the string,
  \begin{equation}
  Tx_\mu' + q_i F_{\mu\nu}(x){\dot{x}}^\nu=0, \quad i=1,2,
  \end{equation}
where $q_1$ and $q_2$ are the charges at the end points. To some extent this is a boundary requirement that parallels the one used in our model, as we also require the positions to be fixed at the ends, at all $t$, as shown in equations (\ref{constraints1}) and (\ref{constraints2}) above. However, there is an important difference: in the electromagnetic case the boundary conditions are {\it dynamic}, implying a balance of elastic and electromagnetic forces at the ends. In our case the conditions are by contrast purely geometric in nature, not referring explicitly to forces acting at the ends.

It is somewhat surprising nevertheless that calculated results in the two cases are in good agreement with each other.  For instance, our zero-point energy (\ref{60}) is precisely the same as Nesterenko's  expression (5.31) for a string moving in an external magnetic field.  Although the physical models are different from each other, this agreement indicates the robustness of the regularization method, especially the one involving use of the Hurwitz zeta function, under different physical conditions.

\appendix
\numberwithin{equation}{section}
\section{Appendix}
To evaluate the infinite sum in the first line of (\ref{56}) we write
{\setlength\arraycolsep{2pt}
\begin{eqnarray}
\nonumber
\sum_{n=-\infty}^\infty\ln(n_r^2+\kappa^2)&=&\ln\prod_{n=-\infty}^\infty(n_r^2+\kappa^2)\\\nonumber
 &=&\ln\bigg[\prod_{n=-\infty}^\infty(n_r+i\kappa)\\
 &&\times\prod_{n=-\infty}^\infty(n_r-i\kappa)\bigg].
\end{eqnarray}}
where $n_r=n+r$. Then by virtue of the formula
\begin{equation}\label{1}
\prod_{n=-\infty}^\infty(nx+y)=\sin\Big(\frac{\pi x}{y}\Big),
\end{equation}
we get
{\setlength\arraycolsep{2pt}
\nonumber
\begin{eqnarray}
\nonumber
\sum_{n=-\infty}^\infty\ln(n_r^2+\kappa^2)
  &=&\ln \sinh\pi(\kappa+ir) \\
  &&+\ln\sinh\pi(\kappa-ir).
\end{eqnarray}}

\noindent {\bf Acknowledgments}

\noindent  I. B. thanks   Vladimir Nesterenko for valuable information. The same author acknowledges financial support from the Research Council of Norway, Project 250346. The work of A. J. has been financially supported by Research Institute for Astronomy \& Astrophysics of Maragha (RIAAM) under project No.1/6025-65.

}

\end{document}